\documentclass[useAMS,usenatbib]{mn2e}
\usepackage{ifpdf} 
\ifpdf
  \pdfoutput=1  
  \usepackage[pdftex]{graphicx}         

  \usepackage[colorlinks, bookmarks, breaklinks, pdftitle={Estimating Black Hole Masses in Triaxial Galaxies},pdfauthor={Remco C. E. van den Bosch and P. Tim de Zeeuw}]{hyperref}
  \hypersetup{linkcolor=black,citecolor=black,filecolor=black,urlcolor=blue}       
  \usepackage[kerning]{microtype}  
\else                                
  \usepackage[dvips]{graphicx} 
\fi

\usepackage{mathptm}
\usepackage{listings}
\usepackage[fleqn]{amsmath}
\usepackage{amssymb}

\usepackage{relsize}  
                   
\newcommand{\plotone}[1]{%
  \includegraphics[width=1.0\columnwidth]{#1}}
\newcommand{\plottwo}[1]{%
  \includegraphics[width=\textwidth]{#1}}
 \newcommand{\plotoneplus}[2]{%
  \includegraphics[width=\columnwidth,#2]{#1}}

\def\kms{km\,s$^{-1}$}
\def\dgr{$^\circ$}
\def\Mbh{\textsc{m}$_\bullet$}
\def\Rbh{\textsc{r}$_\bullet$} 
\def\Re{\textsc{r}$_e$} 
 
\def\Msun{\textsc{m}$_\odot$}

\def\MLsuni{\textsc{m}$_\odot$/\textsc{l}$_{\odot,I}$} 
\def\ML{\textsc{m}\relsize{-1}/\relsize{+1}\textsc{l}}

\def\Msigma{\Mbh-\relsize{-1}$\sigma$\relsize{+1}}

\def\Sauron{\textsc{sauron}}
\def\Oasis{\textsc{oasis}}

\def\Wfpc{\textsc{wfpc}}
\def\Wfpc2{\textsc{wfpc\relsize{-1}2\relsize{+1}}}
\def \Stis{\textsc{stis}} 
\def \M32{\textsc{m}\relsize{-1}32\relsize{+1}}

\def \NGC{\textsc{ngc~}}
\newcommand{\NGCN}[1]{\textsc{ngc}\relsize{-1}~#1\relsize{+1}}

\def\ML{\textsc{m}/\textsc{l}}

\newcommand{\colfigref}[1]{\ref{#1}}

\usepackage{journalnames}

\title[Estimating Black Hole Masses]{Estimating Black Hole Masses in Triaxial Galaxies} 
\author[van den Bosch \& de Zeeuw]{Remco C.~E. van den
Bosch$^{1,2}$, P.~Tim de Zeeuw$^{2,3}$ \\ 
$^1$ McDonald Observatory, The
University of Texas at Austin, 
TX 78712, Austin, USA \ifpdf
\href{mailto:bosch@astro.as.utexas.edu}{    
[bosch@astro.as.utexas.edu]} 
\else
[bosch@astro.as.utexas.edu]
\fi \\
$^2$ Sterrewacht Leiden, Universiteit Leiden, Postbus 9513, 2300 RA Leiden, The Netherlands \\
$^3$ European Southern Observatory, D-85748 Garching bei M\"unchen\\
}
  
\date{Accepted 2009 October 2.  Received 2009 October 2; in original form 2009 February 27}

\pagerange{\pageref{firstpage}--\pageref{lastpage}} \pubyear{2009}
\begin{document}
    
\maketitle

\label{firstpage}

\begin{abstract} 
Most of the super massive black hole mass (\Mbh) estimates based on stellar kinematics use the assumption that galaxies are axisymmetric oblate spheroids or spherical. Here we use fully general triaxial orbit-based models to explore the effect of relaxing the axisymmetric assumption on the previously studied galaxies \M32 and \NGCN{3379}. We find that \M32 can only be modeled accurately using an axisymmetric shape viewed nearly edge-on and our black hole mass estimate is identical to previous studies. When the observed 5\dgr\ kinematical twist is included in our model of \NGCN{3379}, the best shape is mildly triaxial and we find that our best-fitting black hole mass estimate doubles with respect to the axisymmetric model. This particular black hole mass estimate is still within the errors of that of the axisymmetric model and consistent with the \Msigma\ relationship. However, this effect may have a pronounced impact on black hole demography, since roughly a third of the most massive galaxies are strongly triaxial.

\end{abstract}


\section{Introduction}

The masses of super massive black holes in the centers of galaxies are known
to correlate with several properties of the host galaxy. The most well known
correlation is with the stellar velocity dispersion of the galaxy
\citep[\Msigma, e.g.,][]{2002ApJ...574..740T}. The black hole is thought to
play an important role in the evolution of its host (e.g. AGN feedback), as
the properties of galaxies are tightly linked to \Mbh. Therefore it is
important to be able to measure the \Mbh\ accurately and understand whether
the scatter in the relations is due to measurement error, or if it is
intrinsic. Most of the \Mbh\ estimates upon which these relationships are
based were derived using dynamical (edge-on) axisymmetric (or spherical)
dynamical models (See \citealp{2005SSRv..116..523F} for a review).

It has long been known from photometry that some elliptical galaxies are
triaxial \citep[e.g.][]{1996ApJ...464L.119K}, i.e.\ have intrinsic shapes with
three distinct axes \citep{1976MNRAS.177...19B, 1978MNRAS.183..501B}, and more
recently \cite{2007MNRAS.379..401E} and \cite{2007MNRAS.379..418C} have shown
using stellar kinematics that around a third of the most massive ellipticals
are at least mildly triaxial. It is thus relevant to rederive the \Mbh\ in
these galaxies with our triaxial instead of axisymmetric orbit super-position
models.

\cite{2007MNRAS.381.1672T} have modelled mock triaxial merger remnants using axisymmetric geometry and found a correlation between viewing angle and the recovered total galaxy mass. Additionally, they found that the luminous mass to light ratio (\ML) was underestimated by up to 20\% confirming that triaxial modelling is preferred.

In this paper we explore the black hole recovery with the triaxial machinery from \cite{2008MNRAS.385..647V}\ using galaxies that have previously been modeled with axisymmetric codes to verify the triaxial models. We do this with \M32\ and \NGCN{3379} which have their black hole mass determined using \Stis, as well as \Sauron~and \Oasis~Integral Field Unit (IFU) data. Both galaxies have state-of-the-art kinematics available over a large spatial range and are inside the sphere of influence of the black hole.

We describe our modeling technique and uncertainties in \S\ref{method} and
then derive our black hole estimates on galaxies \M32\ and \NGCN{3379}\ in
\S\ref{sec:bh3}. We briefly address the reliability in
\S\ref{sec:tbh_reliability}, and we end with discussion and conclusions in
\S\ref{sec:tbh_discuss}.

\section{Triaxial Schwarzschild modeling} \label{method}

In this paper we use the triaxial \cite{1979ApJ...232..236S} orbit
superposition technique as it is described in \cite{2008MNRAS.385..647V}, of
which we give a brief summary here. It is a powerful tool to construct
realistic dynamical models. It allows for an arbitrary triaxial gravitational
potential (with possible contributions from dark components) in which the
equations of motion are integrated numerically for a representative library of
orbits. Then the superposition of orbits is determined for which the combined
density and velocity moments best fit the observed surface brightness and
kinematics using least squares. By marginalising over parameter space, Schwarzschild's method not
only provides the viewing direction and the \ML\ with the dark matter
contribution, but also allows the investigation of the intrinsic dynamical
structures as well as the distribution function through the orbital mass
weights \citep[cf.][]{1984ApJ...287..475V}. These models have complete
freedom: specifically no form of (an-)isotropy is implied, within the limits
of the observed photometry and stellar kinematics.

\cite{2005MNRAS.357.1113K} showed that the (3I) Schwarzschild method can
recover the phase-space distribution function in the two-integral axisymmetric
case. Additionally, \cite*{2008MNRAS.385..614V} confirmed that the same is
true for triaxial Schwarzschild models and extended this to show that we can
recover the internal dynamics and three-integral distribution function of
triaxial early-type galaxies given their viewing angles. In \cite{2009TRIAXBH}
we have done tests on several three-integral Abel models, that simulate real
early-type galaxies, and found we are also able to recover the viewing angles
of triaxial early-type galaxies, as long as their kinematics show clearly
defined gradients. These final tests allowed us to firmly establish robustness
of the shape recovery and phase space distribution of triaxial Schwarzschild
modeling.

To construct a (luminous) mass model we assume that the three-dimensional mass distribution can be parameterized with multiple coaxial Gaussians \citep{1992A&A...253..366M, 1995AJ....109..572B, 1994A&A...285..723E, 2002MNRAS.333..400C}. The mass distribution used in the models is constructed by fitting two-dimensional Gaussians to the broad band photometry of the galaxy. These Gaussians can be deprojected onto a coaxial triaxial shape by choosing three viewing angles; ($\vartheta$, $\varphi$) and the apparent misalignment ($\psi$), which are used to define the direction from which the galaxy is seen. The shape of each deprojected Gaussian depends on the viewing angles, its observed flattening and its isophotal twist. The shape of each three-dimensional Gaussian is characterized by $p_j=b_j/a_j$, $q_j=c_j/a_j$ and $u_j=a'_j/a_j$, where $a_j$, $b_j$ and $c_j$ are the long, intermediate and short axis lengths of each individual Gauss $j$, and $a'_j$ is the length of longest axis of the Gauss as observed on the sky. To do a full search of all possible mass models we first sample uniformly over a (separate) set of axis ratios, $p$, $q$ and $u$, which translates to a set of viewing angles. These viewing angles are then used to construct all the mass models. The light model is converted to a mass by assuming a constant \ML\ ratio \citep[but see][]{2006ApJ...641..852V, 2006A&A...445..513V}.

The black hole mass estimates determined using Schwarzschild modeling have not been without debate: \cite{2004ApJ...602...66V} reported the existence of a $\chi^2$ plateau prohibiting a \Mbh\ assessment. This could be avoided by using some form of regularisation and enough orbits in the models. The debate was settled by \cite{2006MNRAS.373..425M} who showed that the results from standard superposition methods are accurate, if observational errors are taken into account. Currently, McDermid is leading a joint effort in a comparative study of the \Mbh\ recovery using three independent axisymmetric codes. Using generalised cross validation, he finds that regularisation should be used to avoid the known issue of these models over-fitting the data, and thus yield reliable error estimates on \Mbh. While these results are obtained with axisymmetric modeling, it is likely that they will hold in the triaxial case too, as the triaxial method is conceptually very similar. In this paper we repeated all modeling with and without regularisation and found no difference in the recovered black hole masses.

To determine the uncertainty on the derived shape we shall use the $\chi^2$ based confidence interval that was established in \cite{2009TRIAXBH}. The intervals are based upon the expected standard deviation of the $\Delta\chi^2$(=$\sqrt{2N_{obs}}$, where $N_{obs}$ is the number of kinematical observations used to constrain the model). As the \Mbh\ determination is only influenced by a few of the innermost kinematical observations, we shall use the standard formal 1$\sigma$ and 3$\sigma$ results based upon a $\chi^{2}$ distribution with two degrees of freedom. This is based on the assumption that the determined shape is independent of the \Mbh. It is not obvious that this assumption holds, but as we shall see in the rest of this paper, it works and we discuss the validity of this assumption in \S\ref{sec:tbh_reliability}.

\section{Black hole estimates using triaxial models}
\label{sec:bh3}   

Here we describe the results of our dynamical modelling (as described in
\S\ref{method}) of \M32 and \NGCN{3379}. We compare the near-axisymmetric
\Mbh\ estimates with literature values and explore the effect of the possible
triaxial shape upon the black hole mass estimate and orbital structure.

\subsection{M32} \label{M32} 

To be able to compare our results directly to other studies it was important
that we used galaxies that have a published \Mbh\ obtained using axisymmetric
Schwarzschild models.

Therefore we chose the nearby compact fast rotator E3 galaxy \M32
\citep{1985ApJ...295...73K}, as it has been investigated by several
independent authors \citep[e.g.][]{1998ApJ...493..613V, 2002MNRAS.335..517V,
2004cbhg.symp....1K,2004ApJ...602...66V}. It is consistent with axisymmetry,
as it shows regular rotation and has almost no isophotal twist
\citep[$<3$\dgr,][]{1993A&A...271...51P,1998AJ....116.2263L}. For our modeling
of \M32\ we assumed a distance of 0.79 Mpc and used the surface brightness
distribution, and the wide field \Sauron~data from \cite{2007MNRAS.379..418C}
and the \Stis\ data from \cite{2001ApJ...550..668J}, that probes well within
the sphere of influence of the black hole. The total number of kinematical
observations used to constrain the models is 58 from \Stis\ and 964 from the
\Sauron\ observations, measured up to the Gauss-Hermite moment
\citep{1993ApJ...407..525V, 1993MNRAS.265..213G, 1997ApJ...488..702R} $h_{4}$.
The \Sauron\ kinematics were point-symmetrized as described in Appendix
\ref{pointsym}. Because \M32\ has a dispersion ($\sim$90 \kms) below the
instrumental dispersion of \Sauron\ (120 \kms) the moments $h_{3}$ and higher
are hard to measure, and thus have large associated errors
\citep{2004PASP..116..138C, 2004MNRAS.352..721E}. However, they are still
included in the models to ensure that the (otherwise unconstrained)
reconstructed LOSVDs do not deviate too far from a Gaussian-like shape.

We did not use the high spatial resolution (HR) \Sauron~observations from
\cite{2002MNRAS.335..517V}, as the kinematic extraction of the low resolution
data by \cite{2007MNRAS.379..418C} is superior, due to a much better
extraction method and significantly improved stellar templates. Since we have
high resolution data from \Stis, we chose not to re-reduce the HR data.

\subsubsection{Triaxial models of M32} 
\label{m32shape}

\begin{figure}

\plotone{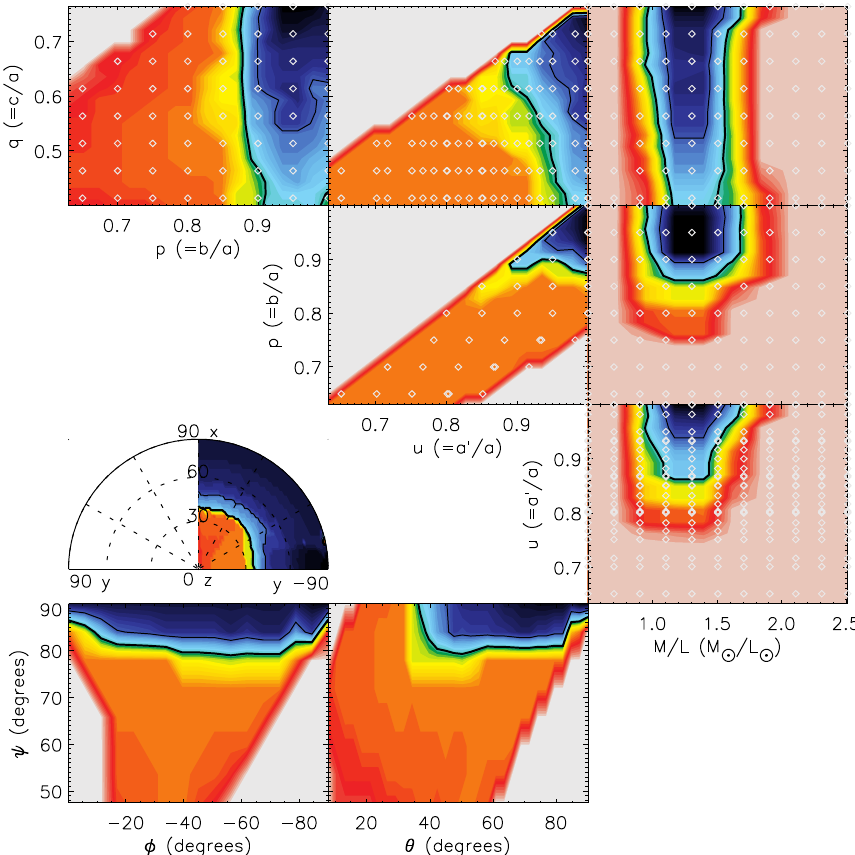}

\caption{Contour maps of the confidence intervals for the parameters of the mass model of the models of M32. From blue to red the color denote high to low confidence. Areas for which the surface brightness cannot be deprojected are white. The contours denote 1, 3 (thick line) sigma confidence levels at $\Delta\chi^2=63$ and 189. The six upper-right panels show the intrinsic shape parameters $(p,q,u)$ and the \ML; the three lower left panels show the viewing angles $(\vartheta,\varphi,\psi)$. The combination of $\vartheta$ and $\varphi$ is shown in a Lambert equal area projection (The half circle), seen down the north pole ($z$-axis). In this panel the $x$, $y$ and $z$ symbols give the location of views down those axis. The best-fitting models are very round and near oblate. The preferred viewing angles are near the y-axis with nearly no intrinsic misalignment. See \S\ref{method} and \S\ref{m32shape}} \label{M32shape}

\end{figure}

Before we can get an \Mbh\ estimate we need to explore the shape of \M32,
because it is too expensive to make models and explore parameter space for the
full range in shape and \Mbh, as it would increase the computation time by a
factor 10. Therefore, we first fix the \Mbh\ at $2.6\times10^{6}$\Msun, while
varying the shape and the \ML. By doing this we assume that the derived shape
does not depend on the specific \Mbh\ we use. We show in \S\ref{m32bhrecov}
and \S\ref{sec:tbh_reliability} that this is a reasonable assumption.

The result of this triaxial modeling is shown in Fig.~\ref{M32shape}. The best-fitting shape is nearly as round as the observed flattening allows and is consistent with an oblate axisymmetric spheroid with axis ratios $(p,q)=(0.95\pm0.05, 0.76^{+0.0}_{-0.2})$. This is fully consistent with expectations given the aligned bi-symmetric rotation in the observed velocity field. Curiously though, \cite{2002MNRAS.335..517V} did find a best-fitting inclination $70^{\circ}\pm10^{\circ}$ (equivalent to $p$=1, $q$=0.73$\pm0.03$) using axisymmetric modeling. Our results do agree, but our errors are different, due to our conservative error bars. Our `axisymmetric' models are still slightly triaxial ($p>0.99$) and thus do allow for additional freedom from the triaxial orbital families. The box orbits are important as our (near) axisymmetric models contain 8\% box orbits within one \Re.

Within the subset of axisymmetric models, the inclination is not well constrained at $\vartheta=i=$\mbox{$90^{\circ}$}$^{+0}_{-50}$. Essentially the full inclination range (given the observed flattening) is allowed at the 3$\sigma$ level. This result is interesting as it agrees with the observation made in \cite{2005MNRAS.357.1113K}, and in \cite{2009TRIAXBH}, that the axisymmetric models cannot constrain the inclination well. However, as is also shown in \cite{2009TRIAXBH}, we can marginalize over the allowed shapes to gain insight into the physical parameters - like \ML\ and anisotropy - which tells us more about the galaxy than the inclination. The \ML\ is $1.4\pm0.2$ \MLsuni, identical to the value from the models of \cite{2007MNRAS.379..418C}\footnote{\cite{2006MNRAS.366.1126C} finds 1.2 \MLsuni.} and \cite{2002MNRAS.335..517V} (when corrected for the different assumed distance and for galactic reddening).

\subsubsection{M32 \Mbh\ estimate}

\label{m32bhrecov} 
\begin{figure}
\plotone{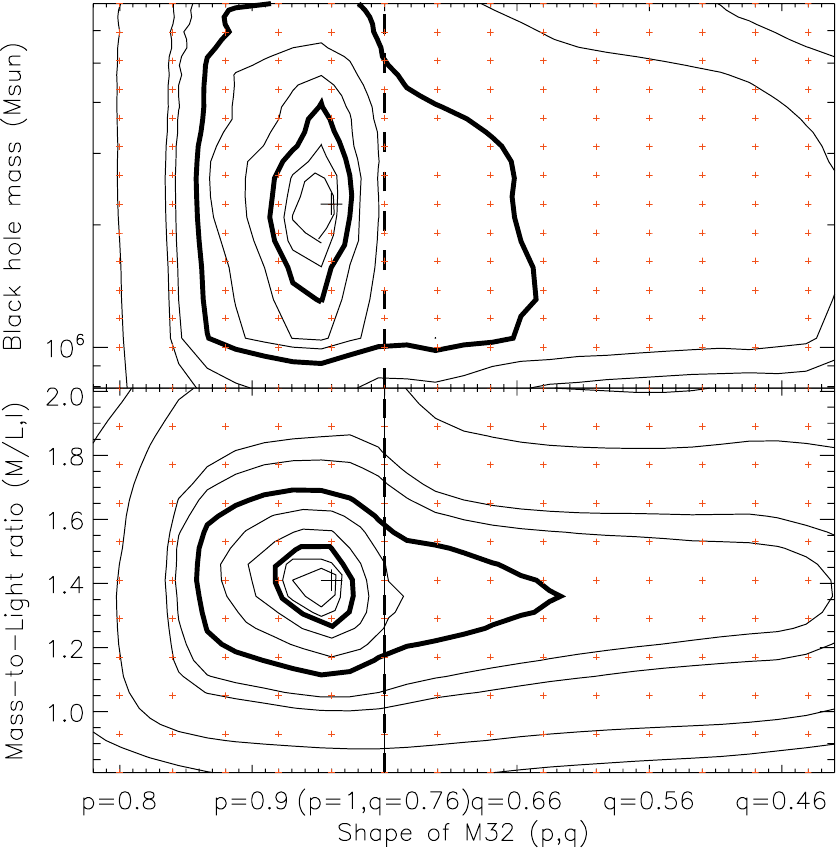}

\caption{\label{M32BH} \Mbh\ (top) and the \ML\ (bottom) confidence levels as a function of shape for \M32. The small red crosses indicate where models were computed and the big black cross indicates the best-fit model. From far left to far right the models change from prolate ($p=0.8$, $q=0.76$), triaxial ($p=0.9$, $q=0.76$) and oblate ($p=1.0$, $q=0.46$). The vertical line denotes the transition from triaxial to oblate. The models on the vertical line are the roundest possible models. The contours show increasing level of confidence. The inner and outer thicrk contour indicate a $\Delta\chi^{2}$ of 14.1 and 63, which correspond to the 99.9\% and 96\% confidence on the \Mbh\ and shape, respectively. The models with \Mbh = 0 are relocated to 8$\times 10^{5}$ \Msun, to place them within the plot.}

\end{figure}

To illustrate the effect of triaxiality, we show what happens when we sample different shapes while varying the \Mbh\ and the \ML. We vary the shape from maximally prolate to maximally oblate (6 models) and at different axisymmetric shapes (8 models). These shapes continuously follow the $q$=0.76 and then the $p$=1 line in the upper leftmost plot ($p$ vs. $q$) in Fig.~\ref{M32shape}. This also fully encompasses the uncertainty in estimated shape.

In Fig.~\ref{M32BH} we show the confidence levels ($\Delta\chi^{2}$)
of the \ML~ and \Mbh\ as a function of these shapes. As expected the
contours show that the roundest models are the best-fit ($q\!>$0.66,
$p\!>$0.88). After marginalizing over the shapes, our \Mbh\ mass
estimate of $(2.4\pm1.0) \times10^{6}$ \Msun is fully consistent
with previous results of $(2-4)\times10^{6}$ \Msun\ from
\cite{2001ApJ...550..668J}, $(2.4\pm 0.7)\times10^{6}$ \Msun\ from
\cite{1998ApJ...493..613V} and $(2.5\pm 0.5)\times10^{6}$ \Msun\
\cite{2002MNRAS.335..517V} and $2.6\times10^{6}$ \Msun\ from \Msigma\
\citep{2002ApJ...574..740T}.

\begin{figure*}
 \plotoneplus{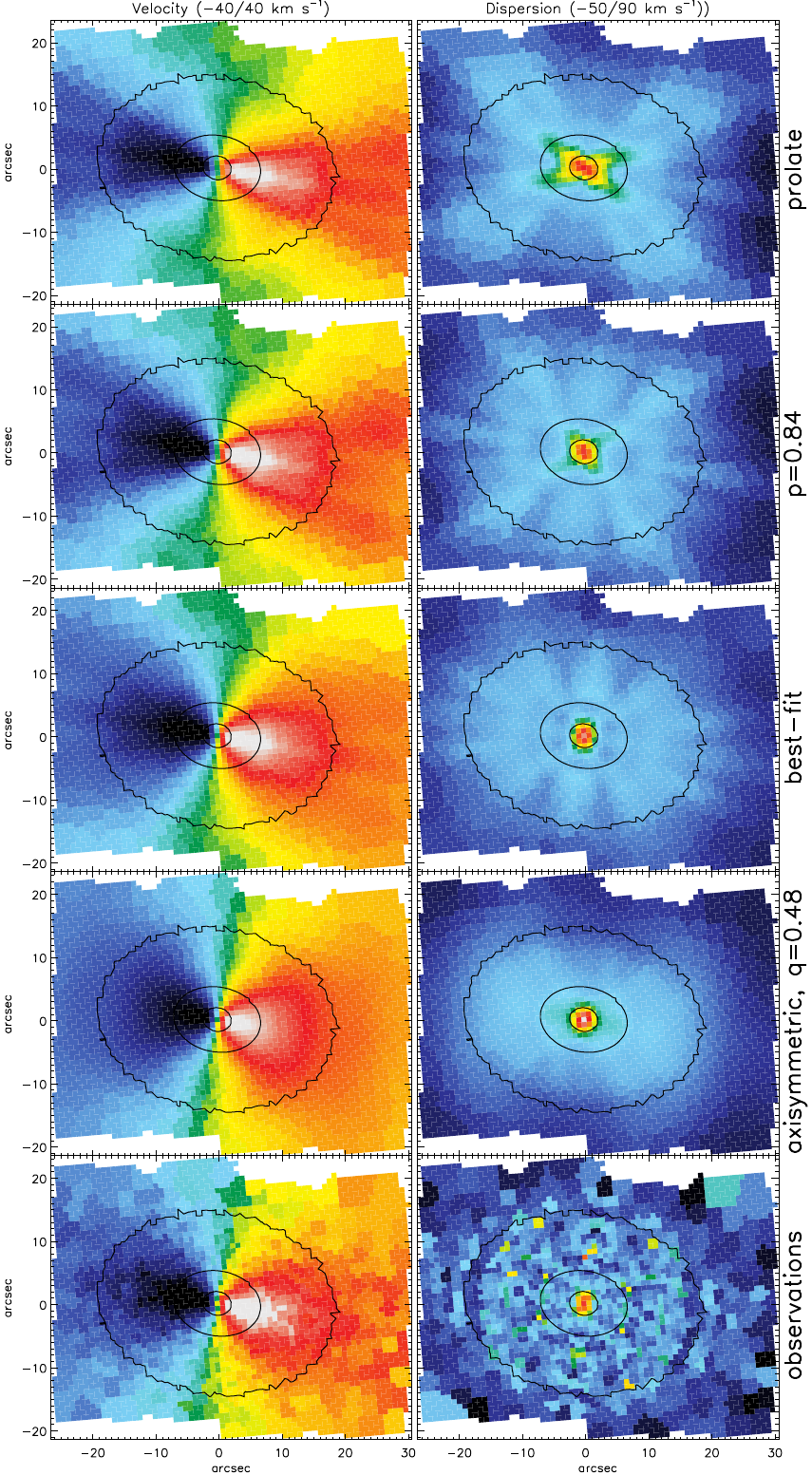}{angle=-90}
 \caption{\label{M32kin}   
 Observed stellar kinematics of \M32\ (left) and the models that correspond to Fig.~\ref{M32BH}, see labels. Top row is the stellar mean velocity and bottom is the stellar velocity dispersion. From left to right: \Sauron\ observations, models from oblate to prolate; ($p,u$) = (1.00,0.48), (0.97,0.76), (0.84,0.76), (0.77,0.76). Contours show representative isophotes. All but the prolate models reproduce the observations well.} 
\end{figure*}

The two most nearly prolate models in this test are a significantly
worse fit than the best-fit model. To illustrate this we show the
kinematics of the observed stellar kinematics and four models, varying
from oblate to prolate, in Fig.~\colfigref{M32kin}. While the oblate and
round models reproduce the observations, the prolate models are unable
to do so, and we can therefore conclude that \M32\ does not have prolate shape.

Interestingly the best-fitting \ML~ and \Mbh\ do not change significantly for the individual triaxial models (See Fig.~\ref{M32BH}), indicating that our initial assumption, i.e. that the recovered shape is not dependent on the initial \Mbh, is probably reasonable in this case. To ensure that our models do not depend on the number of orbits, we doubled the number of orbits, and repeated the above test and found no significant change on the best-fit error and the formal error bars.

Overall, this shows that it is possible to recover \Mbh\ with our triaxial
method. Furthermore, we showed that, in this case, the recovered \Mbh\ does
not significantly depend on the intrinsic shape, and that the shape of \M32\
is very close to oblate.

\subsection{NGC~3379} \label{N3379} 
\defcitealias{2006MNRAS.370..559S}{S06} 

As the second test galaxy we chose the fast rotator \NGCN{3379}, which also has its black hole mass measured with orbit-based models \citetext{\citealp{2000AJ....119.1157G}, Shapiro et al. 2006, hereafter \citetalias{2006MNRAS.370..559S}, \citealp{2007ApJ...664..257D}}, has a decoupled nuclear gas ring \citep[e.g.][]{2001AJ....121..244S} and an \Mbh\ estimate from the gas kinematics \citepalias{2006MNRAS.370..559S}. The velocity maps show regular rotation that is consistent with oblate axisymmetry, and probe to well within the sphere of influence. Detailed axisymmetric orbit-based and gas disc models are shown in \citetalias{2006MNRAS.370..559S}. However this galaxy shows mild evidence for triaxiality: It has a small isophotal twist \citep{1987AJ.....94.1519C}, an $5\pm3$\dgr~kinematical misalignment \citep[\citetalias{2006MNRAS.370..559S},][]{1999AJ....117..126S} and even shows hints of kinematical twist \citep{2008MNRAS.390...93K}. \cite{1991ApJ...371..535C} suggested it might be a triaxial S0 seen face-on, and \cite{2001AJ....121..244S} has argued that this galaxy is mildly triaxial. \cite{2008MNRAS.390...93K} showed that \NGCN{3379} is consistent with being like all the other {\it fast-rotators} in the \Sauron\ sample: nearly axisymmetric and with a disk, but seen almost face-on (based on their V/$\sigma$ diagram). Their interpretation would mean that the misalignment is due to a non-perfect circularity of the disk.

All this makes \NGCN{3379} an ideal test case for the black hole
recovery of the triaxial Schwarzschild machinery. We use the surface
brightness distribution (based upon \Wfpc2 and ground-based imaging),
\Sauron\ and \Oasis\ stellar kinematics, and distance (10.28 Mpc) from \citetalias{2006MNRAS.370..559S} for our modeling.

\subsubsection{\NGCN{3379} in the axisymmetric limit} \label{n3379axi} 

\NGCN{3379} is a very round galaxy, so it is difficult to establish the photometric PA with high precision, and thus the exact amount of misalignment is uncertain ($5\pm3$\dgr), and is in principle consistent with no misalignment (0\dgr) within 2$\sigma$. As such this galaxy could be an axisymmetric oblate spheroid. Exploiting this uncertainty, \citetalias{2006MNRAS.370..559S} aligned the stellar kinematics with the photometry by rotating the velocity field by $5$\dgr\ and bi-symmetrizing the kinematics. This adjustment to the observed kinematics ensures that they are completely compatible with the axisymmetric modelling and these `corrected' observations were used for their \Mbh\ determinations. To check our triaxial modeling machinery, we duplicate those models from \citetalias{2006MNRAS.370..559S} to see if we recover the same \Mbh\ given the same input parameters. This test is important as our triaxial machinery is not capable of making a perfectly axisymmetric model, so even our {\it axisymmetric model} would be minutely triaxial ($p>0.99$) and can still benefit from the additional orbital families, especially the strongly radial box orbits, which do not exist in a pure axisymmetric potential.

\begin{figure}
    \plotone{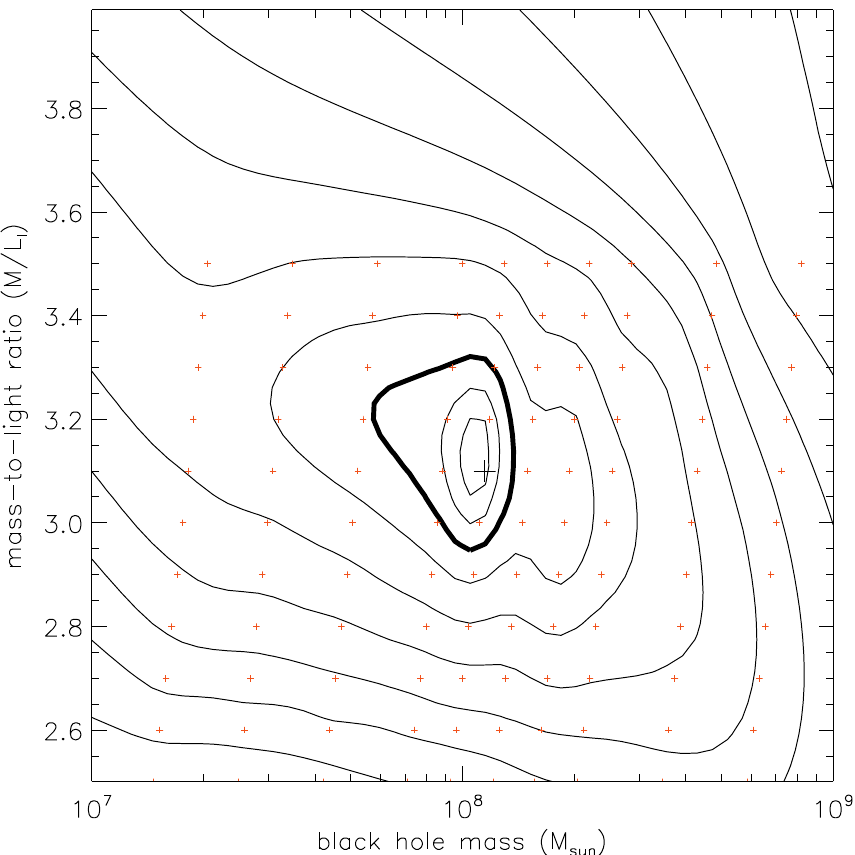} 
    \caption{\label{N3379axi} The confidence levels in the \Mbh~and the \ML~ for an axisymmetric model of \NGCN{3379} made with the triaxial software. See \S\ref{n3379axi}. The thick contour denotes $\Delta\chi^{2}$=14.1, corresponding to a 99.9\% confidence level. The location of the minimum and the shape of the confidence contours match the results from \citetalias{2006MNRAS.370..559S} (see the rightmost panel in their Fig. 8), demonstrating that both codes provide identical results when given the same observables and shape.} 
\end{figure}

We reconstructed the parameter space in \ML\ and \Mbh\ from \citetalias{2006MNRAS.370..559S}. The results, presented in Fig.~\ref{N3379axi}, are identical to the estimate produced with the axisymmetric code, giving \Mbh=$(1.4\pm0.9) \times10^{8}$ \Msun\ and an \ML\ of $(3.1\pm0.2)$ \MLsuni\ and is also similar to the results from \cite{2000AJ....119.1157G}. The fit to the kinematics is not shown, as it is essentially identical to the figures in \citetalias{2006MNRAS.370..559S}. To be completely confident we also tested with double the number of orbits and this does not affect the recovered black hole mass and its confidence interval.

\subsubsection{Triaxial models of \NGCN{3379}} \label{sec:3379triaxmodels} 

\begin{figure*}
\plotoneplus{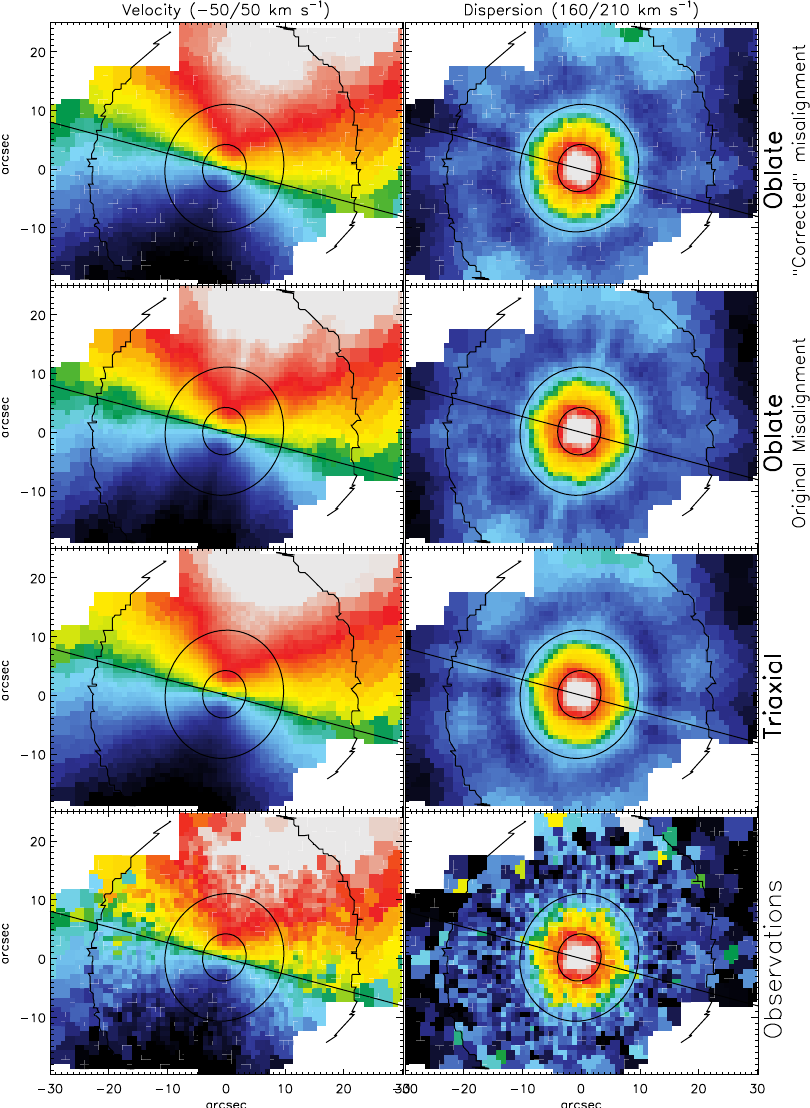}{angle=-90}

\caption{\label{N3379kin} Stellar mean velocity (Top) and velocity dispersion (bottom) of \NGCN{3379}. From left to right: the \Sauron\ observations, best-fitting triaxial model, axisymmetric model and an axisymmetric model without the misalignment. The contours show representative isophotes and the straight line is drawn to guide the eye to the zero-velocity-curve. The triaxial model reproduces the data best, followed by the oblate model and last the oblate model corrected for the misalignment.}

\end{figure*}

\begin{figure}
    \plotone{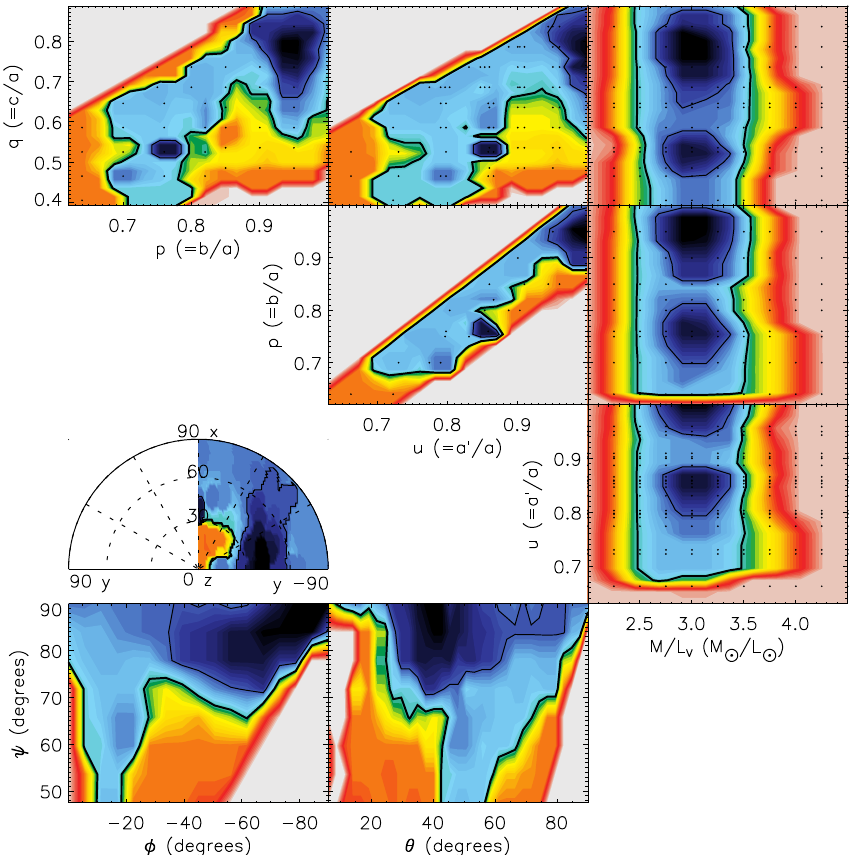} 
    \caption{\label{N3379shape} Recovered shape and viewing angles for the model of \NGCN{3379} with kinematic misalignment. The thin and thick contour represent $\Delta\chi^{2}$=152 and 456, which represent 1 and 3 $\sigma$ confidence intervals. Best-fit model is nearly as round as observed and there is no strong constraint on the viewing angles. Figure layout identical to Fig.~\ref{M32shape}. See \S\ref{method} and \S\ref{sec:3379triaxmodels}} 
\end{figure}

To study the effect of kinematical misalignment on the \Mbh\ estimate we modelled \NGCN{3379} using our triaxial machinery and with the kinematical misalignment intact and point-symmetrized (See Appendix \ref{pointsym}) kinematics. We first searched through the shape distribution, while keeping \Mbh~\ fixed at $1.4 \times10^{8}$ \Msun. The best-fitting kinematics is shown in Fig.~\ref{N3379kin} and \ref{N3379oasis}. The recovered shape, shown in Fig.~\ref{N3379shape} is $(p,q)=(0.95^{+0.03}_{-0.05}, 0.81^{+0.05}_{-0.10})$ at 1 $R_{e}$. Our shape is consistent with the result from \cite{2001AJ....121..244S} of $p \gtrsim 0.99$ and $q\sim0.7$. Given that his method is completely different, it is reassuring to see that we can even reproduce the shape of the confidence intervals \citep[but see][]{2008MNRAS.385..647V}. The allowed viewing angles cover a large range, but prefer strongly inclined (face-on) views, which is also consistent with the results from \cite{2009MNRAS.395...76D}. 

It is important to notice that at the $3\sigma$ level the shape is not constrained well, allowing almost all viewing angles ($\vartheta,\varphi$) and a large allowed range in shapes (see Fig.~\ref{N3379shape}). A pure oblate axisymmetric spheroid is excluded at the $2\sigma$ level, and this happens because the axisymmetric model cannot reproduce the twist in the zero velocity curve ($\Delta\chi^2>200$). This is shown in Fig.~\colfigref{N3379kin}. The differences between the axisymmetric (third panel from the left in Fig.~\colfigref{N3379kin}) and triaxial (second panel from the left in Fig.~\colfigref{N3379kin}) model are not very prominent; the most visible change can be seen in the (twisting) shape of the zero velocity curve.

Since \citetalias{2006MNRAS.370..559S} uses bi-symmetrized kinematics and the axisymmetric models use different intrinsic mass bins it is not possible to directly compare the $\chi^2$ of those models with ours. To do a direct comparison we recreated the original axisymmetric model from \citetalias{2006MNRAS.370..559S} with the triaxial machinery, without bi-symmetrizing, but with the kinematic misalignment correction (see \S\ref{n3379axi}). We expected that this \emph{original} axisymmetric model would fit the data better than the axisymmetric model -- because the latter does not correct for the misalignment -- but this is not what we found (rightmost panel in Fig.~\colfigref{N3379kin}). The kinematics of the best-fit triaxial axisymmetric model are statistically a significantly much better fit, with a difference in $\Delta\chi^{2} > 900$. The differences show up in the twist of the zero velocity curve and the `hexagonal shape' of the velocity dispersion. It seems that in this case the twist of this galaxy was overestimated due to the inaccurate determination of the photometric or kinematic PA. Both are possible because \NGCN{3379} is very round, has some isophotal twist and does not have a strong velocity field. Luckily, the triaxial modeling (as opposed to the axisymmetric modeling) does not depend on these measurements, as it only requires that the relative orientation between the photometry and kinematics be known.
 
Given that the shape can not be constrained accurately our choice of the \Mbh\ might influence the recovered shape. To test if this was the case we checked to see if the recovered shape would differ if we set \Mbh$=0$. The shape recovered was not different.  
 
The intrinsic orientation of the central gas and dust disc in this galaxy is interesting due to its apparent misalignment of $\sim$50 degrees with the main body of the galaxy. The only stable configurations in a stationary triaxial geometry are in the principal planes. \cite{2001AJ....121..244S} did a thorough analysis and concluded that, if the disc lies in a plane, the only option that he could not rule out was a polar ring. \cite{2005AJ....129.2138L} showed that stellar photometry inside 1 arcsecond has a sudden PA twist of more then 20 degrees, towards the gas disc. The gas disc has a size of 4 arcsec and is thus at larger radii than this stellar feature, but the two could be connected. Also, \citetalias{2006MNRAS.370..559S} found evidence that the gas disc might be warped using an ad-hoc model, indicating that a simple stable gas disc in a plane might actually not be a good description. Our current mass model does not include any isophotal twist and therefore does not predict if these two features are intrinsically aligned. To do that, a mass model with isophotal twist would be needed. As an added benefit such a mass model with isophotal twist will lower the amount of possible deprojection, which would indirectly help constrain the shape of this galaxy. 

In our current modelling without isophotal twist, the allowed range in viewing angles is large, and it might thus be possible to place the ring in a principal plane. The ring is misaligned 45\dgr\ from the photometric PA, so to place the ring in a principal plane we need a misalignment of the PA of 45\dgr. Our allowed models do include these extreme misalignments of $\psi=45$ at the 3$\sigma$ level (Fig.~\ref{N3379shape}), but the other viewing angles are then quite restricted: at $\psi=45$ only $(40$\dgr$\!<\!\vartheta\!<\!60$\dgr$,-10$\dgr$\!<\!\varphi\!<\!20$\dgr$)$ are within the 3$\sigma$ contour, essentially disallowing the disc in either the $x-y$ or $y-z$ plane. From our modelling the polar ring is the only possibility, as the inclination of our best-fit models lies below $\vartheta\!<$43\dgr, which is exactly what is needed for the polar ring according to \cite{2001AJ....121..244S}.

\subsubsection{The black hole in \NGCN{3379}} 

\begin{figure}             
   \plotone{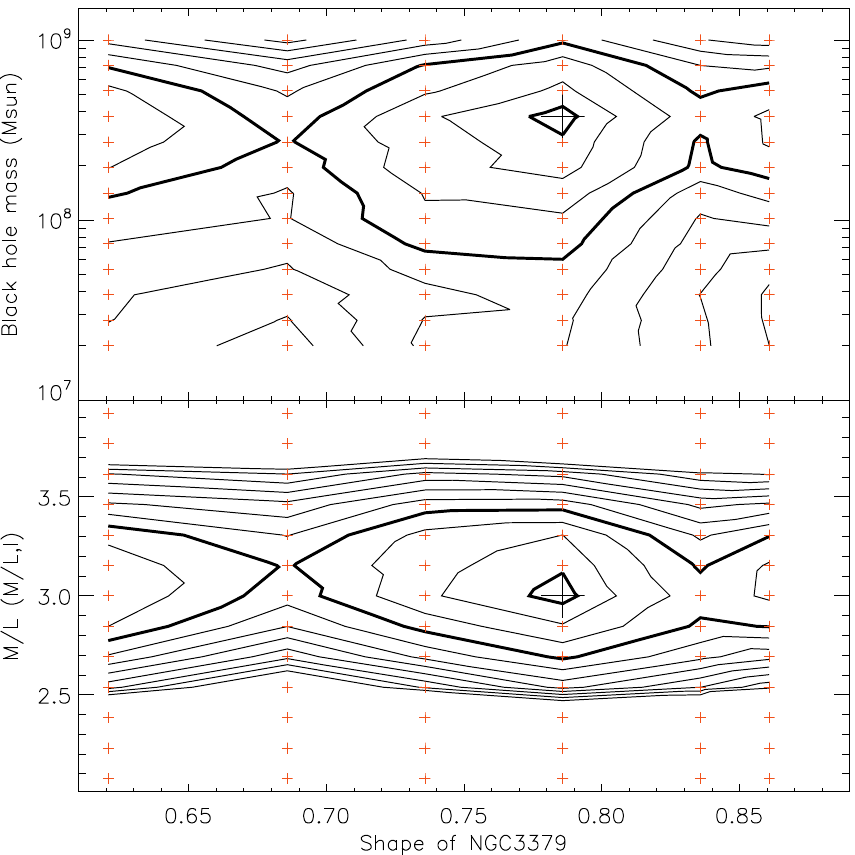}

\caption{\label{N3379bh} The \Mbh\ (top) and the \ML\ (bottom) confidence levels as function of the different triaxial shapes (see text). The small red crosses represent the values for which models where computed and the big black crosses indicate the best-fit model, respectively. The contours show increasing level of confidence. The inner contour indicates a $\Delta\chi^{2}$ of 42.3, further contours represent integer steps of 152, which correspond to the 1,2,3 (thick),4,... $\sigma$ confidence intervals on the shape.}

\end{figure}

Now that we have a handle on the shape, we investigate whether the inferred
shape affects the recovered \Mbh. We used the six best-fitting mass models,
while changing the \Mbh~and the \ML. The results are shown in
Fig.~\ref{N3379bh}. In this figure the shape is parameterized as
$(0.95-p)/2+q$, which is completely arbitrary, but allows us to plot
two-dimensional contours and show how the $\chi^2$ minimum is bracketed. The best-fit shape is independent of the chosen \Mbh, showing that the recovered shape does not depend on the fixed black hole mass that was used in the previous section.  

\begin{figure*}
    \includegraphics{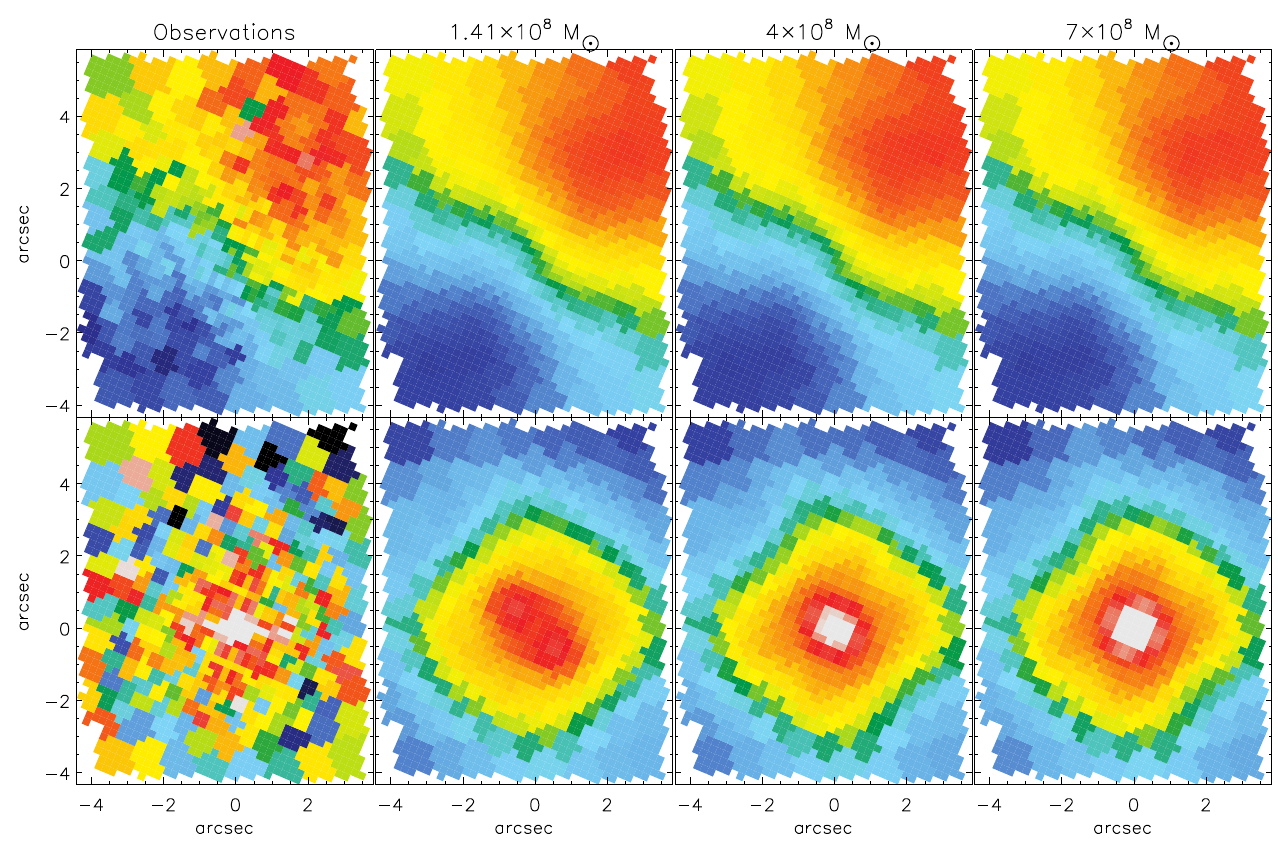} 
    \caption{\label{N3379oasis} Comparison of the \Oasis\ central point symmetrized stellar kinematics (left) of NGC~3379 and models with different black hole masses (1,4,7)$\times 10^8$ \Msun\ on the right. Stellar mean velocity (Top row) ranges from -60...60 \kms and the dispersion ranges from 190...230 \kms. The $4\times 10^8$ \Msun\ is the best-fit model, while the others are not capable of reproducing the observed central dispersion. The twist in the mean velocity field is reproduced perfectly in the triaxial model, which is something that would be impossible for a pure axisymmetric model. While the higher moments are fitted they are not shown as their contribution is not nearly as important as the first two moments (They are shown in S06).} 
\end{figure*}

The best fitting \Mbh\ is $(4\pm1)\times10^{8}$ \Msun\ and the \ML\ is
$(3.0\pm0.2)$ \MLsuni. Surprisingly, this \Mbh\ estimate from the triaxial
model is more than twice as large as $(1.4^{+2.6}_{-1.0})\times10^{8}$ from
the edge-on axisymmetric estimate from \citetalias{2006MNRAS.370..559S}. To
show the quality of the models we show the \Oasis\ kinematics and models with
different \Mbh\ in Fig.~\ref{N3379oasis}. For all the different black hole
masses, the mean velocity field is reproduced extremely well, but the
dispersion is only properly reproduced by the $4\times10^{8}$\Msun\
\Mbh~model. Our result is also significantly above the axisymmetric (near)
face-on model of \citet[$2.0\times10^{8}$ \Msun]{2000AJ....119.1157G} and
\citetalias{2006MNRAS.370..559S}. Our \Mbh~is just outside of the scatter of
the \Msigma\ relationship, as that predicts
$10^{8.13\pm0.06\pm0.27}=(1.4^{+1.5}_{-0.7})\times 10^{8}$ \Msun, placing our
estimate on the heavy side of this relationship.
\citetalias{2006MNRAS.370..559S} showed that the gas disc inside cannot be fit
by simple Keplerian motion, implying that the gas disk is disturbed and may
not be a good candidate for probing the dynamical BH mass. Nevertheless they
constructed an ad hoc and non-unique model and estimated ($2.0\pm 0.1) \times
10^{8}$ \Msun, which is lower than our estimate.


\begin{figure}
    \plotone{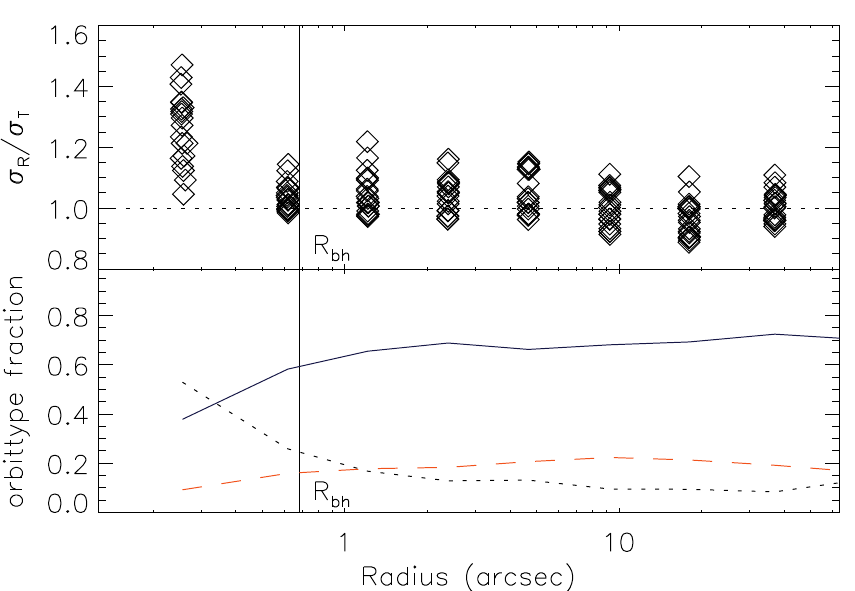}
    \caption{\label{fig:N3379structure} Orbital structure of \NGC3379. Top plot shows orbital anisotropy $\sigma_R/\sigma_T=\sqrt{2\sigma^2_R/( \sigma^2_\phi+\sigma^2_\theta)}$ and the bottom plot shows the orbit type as a function of radius. The model is mildly radially anisotropic outside \Rbh\ and strongly anisotropic inside \Rbh. The short axis tubes (blue solid line) dominate the galaxy outside \Rbh, while the box orbits (black dotted line) become more important inside \Rbh. The long axis tubes (red dashed line) are roughly constant at 20\%.}
\end{figure}

Even with a mildly triaxial shape, the long axis and box orbits can contribute a significant fraction \citep{1992ApJ...389...79H}. Inside one intrinsic \Re\ our model consists of 70\%, 20\% and 10\% short axis, long axis tubes and box orbits respectively. The orbital structure (Fig.~\ref{fig:N3379structure}) of the triaxial model reveals that \NGCN{3379} is radially anisotropic inside the sphere of influence of the black hole (\Rbh) and at most mildly radially anisotropic outside. This is different from \citetalias{2006MNRAS.370..559S}, which showed \NGCN{3379} to be isotropic inside the core radius. In our model, the box orbits contribute most of the mass inside \Rbh, and thus the model becomes strongly radially anisotropic in the center (Fig.~\ref{fig:N3379structure}). This is very different from an axisymmetric model, in which box orbits cannot exist.

The box orbits in the center could even be the cause of our high \Mbh. In the face-on view of these models (28\dgr  $< \vartheta  <$ 49\dgr) the stars on box orbits in the center have the highest dispersion in the direction perpendicular to the viewer and can therefore affect the central observed dispersion. This is exactly opposite to a mechanism suggested by \cite{1988MNRAS.232P..13G} that essentially makes the black hole unnecessary by viewing the galaxy down the $x$-axis (end-on) -- the box orbits would then account for the high velocity dispersion in the center\footnote{This is also in contradiction with our end-on model of \M32, which did need a black hole.}.

\section{Reliability of the black hole mass estimates} 
\label{sec:tbh_reliability}

To get to our best fitting models of \M32\ and \NGCN{3379}, we had to first assume a \Mbh, find the best-fitting shape and then find the best-fitting \Mbh, because the alternative -- searching the full parameter space -- is computationally unpractical. We search the shape parameter space first because an initial guess for \Mbh\ can be done using \Msigma\ and we know from \cite{2009TRIAXBH} that the influence of shape on the quality of the fit is much bigger than that of the black hole, usually more than a factor ten in $\Delta\chi^2$. However, this procedure is not guaranteed to find the global minimum.

To ensure that we do find the minimum, we marginalized over all the best-fitting shapes when searching for the \Mbh. This showed that for both galaxies we found that the best-fitting shape was unchanged by the improved \Mbh\ and that even if we would have chosen a different \Mbh\ beforehand we would still have found the same minimum. Surprisingly, this was also true for \NGCN{3379}, of which the shape is not constrained well and the \Mbh\ estimate changes. This is evidence that the recovery of the intrinsic shape and \Mbh\ are independent, which simplifies future modeling. Figs.~\ref{M32BH} and \ref{N3379bh} also showed the reverse: that the \Mbh\ estimate does not depend significantly on the shape and thus that our \Mbh\ estimate is reliable, even if we do not get the intrinsic shape perfectly correct. The addition of a dark halo in the models would be the next step. \cite{2009ApJ...700.1690G} showed that adding a dark halo can increase the \Mbh\ if the constant \ML\ is overestimated. However this is unlikely to happen for \NGC3379 as the best-fit dynamical \ML\ is already very close to the \ML\ derived from the stellar population \citep{2006MNRAS.366.1126C}. The amount of dark matter in this galaxy has recently been constrained by \cite{2009arXiv0906.0018W}, who measured stellar absorption line kinematics at large radii. By combining both this outer data plus the \Oasis\ observations the effect of the dark matter  on the black hole mass estimate could now be determined.

The final test would be to model an analytical triaxial test galaxy with a
realistic density profile and with a central black hole. However, we are not
aware of the existence of such self-consistent models\footnote{It is possible
to construct analytic triaxial galaxy models with a central black hole and an
$f(E)$ distribution function, but the isodensity surfaces are then identical
to the equipotentials, so that the model would be non-consistent. The value of
using such models as test cases is then limited as we do not expect this
arrangement of isodensity surfaces and equipotentials to arise in real
spheroids, which moreover have anisotropic velocity distributions.}. As an
alternative, galaxies generated using N-body simulations could be used
\citep[similar to][]{2007MNRAS.381.1672T}. However we shall refrain from doing
this test now, as we expect that it would not change the results for the two
nearly oblate axisymmetric galaxies discussed here.


\section{Discussion and conclusions} \label{sec:tbh_discuss} 

We have shown two applications of the triaxial orbit super-position method to galaxies that had their central black hole mass measured with axisymmetric models. The reason for this was two-fold: First, we confirmed that we obtain the same \Mbh\ when using our new triaxial implementation in the oblate axisymmetric limit, confirming that our method is reliable. Secondly, we explored what might happen when the assumption of axisymmetry is relaxed.

For the nearby elliptical galaxy \M32\ we obtained identical results to previous axisymmetric modeling finding a \Mbh\ of ($2.4\pm1.0) \times10^{6}$ \Msun. Our best-fitting shape of \M32\ is very close to oblate axisymmetric, excluding strongly triaxial shapes. The viewing angles are not well constrained.

We also duplicated the models of \NGCN{3379}. After confirming the results from \citetalias{2006MNRAS.370..559S} in the axisymmetric limit, we expanded our search to include triaxiality. While the shape is not strongly constrained, we found that this galaxy is seen almost face-on and can best be described with a triaxial model. The best-fitting models are very round in the center and become fairly flattened at larger radii. These results confirm the claims about the intrinsic shape from \cite{1991ApJ...371..535C}, \cite{1994AJ....108..111S} and \cite{2008MNRAS.390...93K}.


The black hole in this triaxial model weighs $(4\pm 1) \times 10^{8}$\Msun,
which is two times bigger than the axisymmetric estimate from
\citetalias{2006MNRAS.370..559S} and \cite{2000AJ....119.1157G}. We speculate
that the difference in the estimate is due to the combined effect of changing
from an edge-on to a nearly face-on view and the inclusion of the strongly
radial box orbits in the modeling.

The significance of the change of the \Mbh\ of \NGCN{3379} is unclear. While
this individual result is still within the scatter of the \Msigma\ relation,
it can greatly affect all empirical relations based on \Mbh, like \Msigma, if
similar effects are seen in more galaxies. If many intrinsically triaxial
galaxies would have their black hole mass underestimated using axisymmetric
modeling, the empirical relations would be systematically underestimating
black hole masses at the top end, as we believe that triaxial galaxies
dominate at the high mass end \citep[e.g.][]{1996ApJ...464L.119K,
2007MNRAS.379..401E}. To study this in more detail, it is necessary to study
the nuclei of the most massive galaxies with triaxial models.

Also, the axisymmetric models and their \Mbh\ estimates have only been tested
with purely axisymmetric (and spherical) test models. This means that if
axisymmetric models cannot accurately recover black hole masses in triaxial
galaxies, and if most galaxies are significantly triaxial, then this would also
have an impact on black hole demography\footnote{Of course, a similar
argument could be held for the triaxial models, if the modeled galaxies cannot
be described by stable triaxial geometries.}. A strong hint in this direction
is given by \cite{2007MNRAS.381.1672T}, who showed that the axisymmetric
models have difficulty estimating the \ML\ in triaxial galaxies. Based on
their results and our \NGCN{3379} model, we speculate that the \Mbh\ estimates
in triaxial galaxies derived using axisymmetric modeling will have systematic
errors, because the recovery of the \Mbh\ is strongly linked to the \ML.

\section*{Acknowledgements}

It is a pleasure to thank Michele Cappellari, Eric Emsellem, Davor
Krajnovi\'c, Richard McDermid, Glenn van de Ven and Anne-Marie Weijmans for
stimulating comments and lively discussions. We would also like to thank
Kristen Shapiro and Richard McDermid for the reduced \Oasis\ kinematics, and
Eric Emsellem and Michele Cappellari for the \M32\ \Sauron\ kinematics. This
project is made possible through grant 614.000.301 from NWO, Leids
Kerkhoven-Bosscha Fonds and the Netherlands Research School for Astronomy
NOVA. Part of this work is based on data obtained from the ESO/ST-ECF Science
Archive Facility, the Canadian Astronomy Data Centre operated by the National
Research Council of Canada with the support of the Canadian Space Agency, the
NASA/IPAC Extragalactic Database (NED) which is operated by the Jet Propulsion
Laboratory, California Institute of Technology, under contract with the
National Aeronautics and Space Administration and the 1.3m McGraw-Hill
Telescope of the MDM Observatory. The \Sauron\ observations were obtained at
the William Herschel Telescope, operated by the Isaac Newton Group in the
Spanish Observatorio del Roque de los Muchachos of the Instituto de
Astrof\'{\i}sica de Canarias. Most of the models presented in this paper were
computed at the Texas Advanced Computing Center (TACC) at The University of
Texas at Austin.



\appendix

\section{Point symmetrizing velocity maps} \label{pointsym}

\begin{figure*} \plottwo{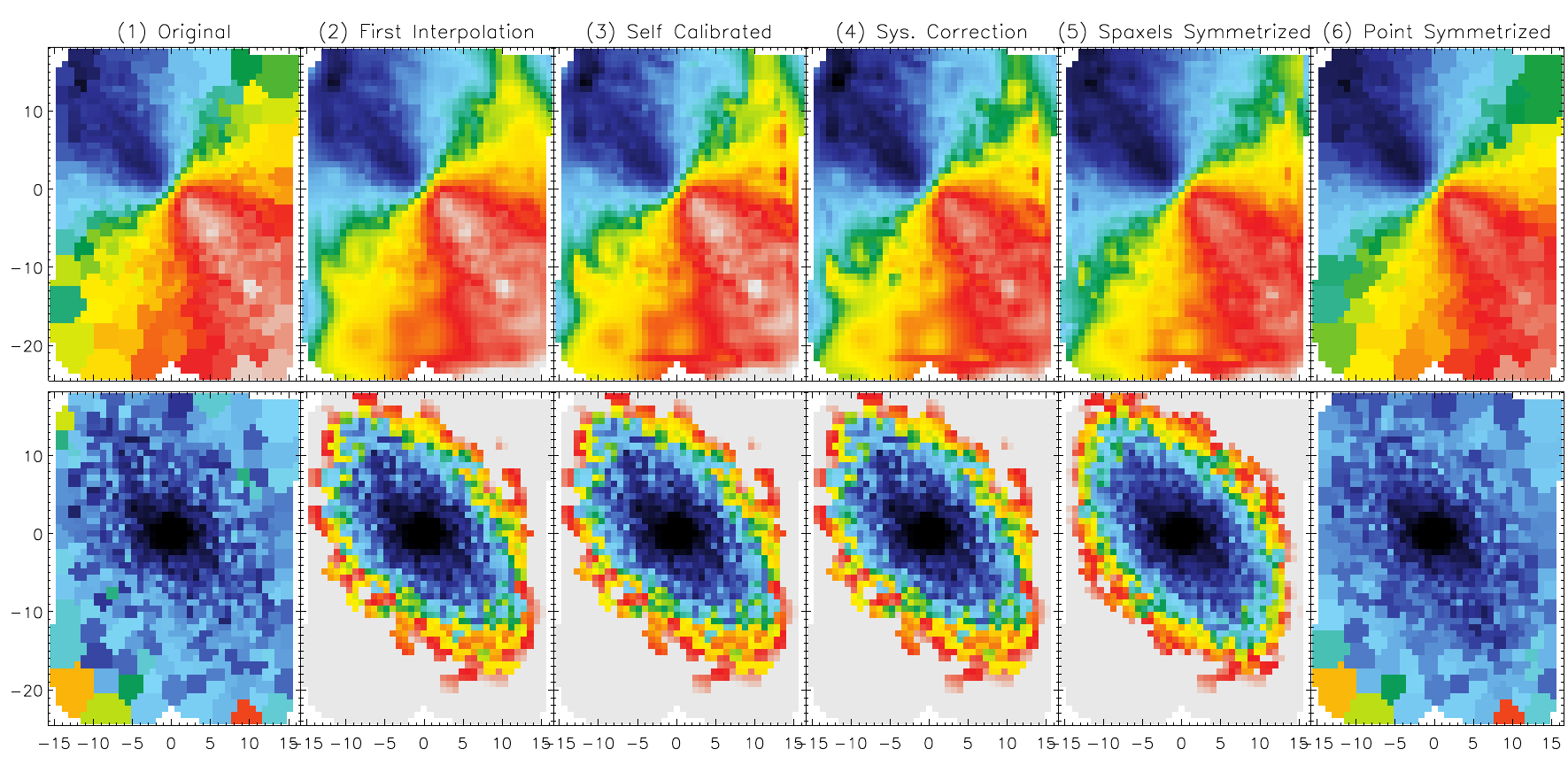} \caption{\label{figsymm} Example
of the point symmetrization process on the \Sauron\ observations of NGC~3377
from Emsellem et al (2004). The top panels show the mean stellar velocity
(-110...+110 \kms) and the bottom row shows the error (2...20 \kms).
\textbf{(1)} The original observations. \textbf{(2)} The errors after being
expanded onto the spaxels and the kinematics after the first interpolation
step. \textbf{(3)} After the self calibrating interpolation steps.
\textbf{(4)} After the velocity field is corrected for the systematics
offsets. \textbf{(5)} After the spaxels are point symmetrized. \textbf{(6)}
After the final step of binning together the spaxels back into bins. Notice
how the errors in the bottom region have not changed after symmetrization,
because those bins do not have a point-symmetric counterpart.
}\nocite{2004MNRAS.352..721E} \end{figure*}

Almost all orbit-based modellers usually symmetrize their input kinematics \citep[e.g.][]{2003ApJ...583...92G, 2006MNRAS.366.1126C}, as their model assumptions enforce bi- or point symmetry anyway. There are numerous reasons to do this. Most commonly this is done to reduce the noise in the observations, facilitate $\chi$-by-eye comparison, force the observations to be bi- or point symmetric, or because they want to reduce the number of observables. It is also a good method to remove systematic effects, including systematic offsets in the odd moments. The symmetrization also improves the linear Gauss-Hermite moments reconstruction used in our models \citep{1997ApJ...488..702R}.

Symmetrizing velocity maps is a degenerate problem and there are an infinite number of \emph{solutions} and none are perfect. In this appendix we describe a novel method to point-symmetrize \Sauron\ velocity fields. It is accurate, conserves the amount of spatial information and propagates the errors, without making unnecessary assumptions. The IDL-script itself is available in the electronic tarball on arxiv.org. At the very least, this method is useful to determine the systematic offsets in the odd velocity moments. The multi-step process is depicted in Fig.~\ref{figsymm}.

We make three assumptions: First we assume that the Gauss-Hermite moments are orthogonal and un-correlated. This assumption is also enforced by the dynamical model itself and is therefore `fair', but not true. Second, we assume that the velocity field varies linearly along spatial coordinates, which is generally true to first order. Third, we assume that we can safely symmetrize without worrying about the PSF.

The \Sauron\ kinematics are typically binned \citep{2003MNRAS.342..345C}. This
means that most kinematic observations (and associated error) span several
spaxels (i.e. lenslets). But still every spaxel has an individual flux and
Signal-to-Noise measurement. We assign every spaxel an error based on its
relative flux\footnote{It is also possible to adapt S/N
instead of flux.} within that bin and the error of the kinematic observation
($\Delta_{kin}$) of that bin as follows.

\begin{equation} \label{spaxerr}
  \Delta^{spaxel}_{i}=\Delta_{kin} \sqrt{\Sigma^{bin}/\Sigma^{spaxel}_{i} }  
\end{equation} 

\noindent Where $\Sigma^{bin}$ is the total flux in that bin and
$\Sigma^{spaxel}_i$ is the flux in the spaxel $i$ (\textbf{2} in
Fig.~\ref{figsymm}). This definition conserves the error of the original bin
when the spaxels are combined back into a bin later. The spaxels can be
recombined into the bins using the weighted mean:

\begin{equation} \label{binrecomb}
    V^{bin} = \frac{\sum_{i} V^{spaxel}_{i}\Delta^{spaxel}_{i}}
           {\sum_{i}\Delta^{spaxel}_{i}}
\end{equation}

\noindent $V^{spaxel}_{i}$ are the kinematics of each individual spaxel, which are yet unknown. To estimate them we construct a linear interpolation of the velocity field over the individual spaxels, using the bin centers as the nodes in the linear interpolation\footnote{The exact way in which this interpolation is done is not important. We use the default IDL interpolater TRIGRID, which does a decent job of extrapolating too.} (\textbf{2} in Fig.~\ref{figsymm}). To be as conservative as possible we need to make sure that after this linear interpolation the combined spaxels still reproduce the original kinematics. Using eqn.~\ref{binrecomb} we compute for each bin what the difference is between the original kinematics and the interpolated one. This difference $D_{bin}$ is then used to self-calibrate the linear interpolation step, by adjusting the velocities at the nodes using $D_{bin}$ and recomputing the interpolation and this is repeated as needed. This self-calibration typically converges quickly and ensures that the velocity map interpolated over the spaxels reproduces the original binned velocity map when binned back (\textbf{3} in Fig.~\ref{figsymm}).

In a triaxial stellar system the odd kinematic moments have to be
anti-correlated across the center. The \Sauron\ reduction pipeline
centers the central spaxel on the galaxy center. Therefore it is trivial to
find the point symmetric counterpart of each spaxel. We can now use this
information to correct for systematic offsets (e.g. recession velocity and
template mismatch) in the velocity map of the odd moments. It is
important to do this before symmetrizing, because bins that do not have a
point symmetric counterpart will otherwise not get corrected for this
offset. For each spaxel that has a counterpart, we compute their weighted
mean velocity and combined error. We then take those values plus the value of
the central spaxel and compute the total weighted average, which is the
systematic offset. We add this systematic offset to the velocities of the
individual spaxels (\textbf{4} in Fig.~\ref{figsymm}).

Now we do the actual point-symmetrizing of the spaxels. This step is almost identical to the previous step, but has some significant differences. For each spaxel that has a counterpart, we compute their weighted mean velocity and combined error. We then replace the values of those spaxels with the weighted mean velocity that we just computed and replace their errors with the combined error multiplied by $\sqrt2$. The error is corrected, because we just stored the velocity in two spaxels and we do not want to artificially decrease our errors (\textbf{5} in Fig.~\ref{figsymm}). After we have point-symmetrized all the spaxels (once) in this way, we combine the spaxels (and their errors) back into the original bins. Now the point symmetrization is complete (\textbf{6} in Fig.~\ref{figsymm}). The end result has exactly the same bins as the original observations and has thus conserved the spatial information.

In principle it is possible to adapt this routine to bi-symmetrization too. However, in the bi-symmetric case no unique counterpart of the spaxels exist, because the position angle (PA) of the (presumed axisymmetric) galaxy does not have to coincide with the pixel grid. To extend our algorithm to bi-symmetrization another interpolation or supersampling needs to be added. Alternatively, appendix C in \cite{2006MNRAS.366..787K} describes a way to bi-symmetrize velocity fields, which interpolates the kinematics between the bin centroids, without self-calibrating, using an un-weighted mean and without propagating the errors. Bi-symmetrization also critically depends on the existence and determination of the global PA. Since we model (axisymmetric) galaxies here with our triaxial method, we can avoid these issues by applying point-symmeterization instead of bi-symmeterization.

The method described here is not perfect and no method for velocity map symmetrizing ever will be. In principle it should be possible to relax some of our assumptions and design a better algorithm. Instead, it would be much more useful to update the orbit-based models, so that they are robust against the systematics in the unsymmetrized kinematics.



\bsp 

\label{lastpage}
\end{document}